\documentclass[pre,aps,twocolumn]{revtex4}

\usepackage{graphicx}
\usepackage{amsmath,empheq}
\usepackage{amssymb}
\usepackage{ifthen}
\usepackage{booktabs}
\usepackage{graphicx}
\usepackage{dcolumn}
\usepackage{bm}
\usepackage{longtable}
\usepackage{gensymb}

\usepackage{graphicx}
\usepackage{dcolumn}
\usepackage{bm}
\usepackage{longtable}
\usepackage{gensymb}
\usepackage{color}

\newcommand{\bb}{\begin{equation}}
\newcommand{\ee}{\end{equation}}
\newcommand{\ba}{\begin{eqnarray*}}
\newcommand{\ea}{\end{eqnarray*}}
\newcommand{\rhor}{\rho({\bf r})}
\newcommand{\dd}{{\rm d}}
\newcommand{\rr}{{\mathbf r}}
\newcommand{\dr}{{\rm d}{\bf r}}

\begin{document}

\title{Critical effects and scaling at meniscus osculation transitions}

\author{Andrew O. \surname{Parry}}
\affiliation{Department of Mathematics, Imperial College London, London SW7 2BZ, UK}
\author{Martin \surname{Posp\'\i\v sil}}
\affiliation{
{Department of Physical Chemistry, University of Chemical Technology Prague, Praha 6, 166 28, Czech Republic;}\\
 {The Czech Academy of Sciences, Institute of Chemical Process Fundamentals,  Department of Molecular Modelling, 165 02 Prague, Czech Republic}}

\author{Alexandr \surname{Malijevsk\'y}}
\affiliation{ {Department of Physical Chemistry, University of Chemical Technology Prague, Praha 6, 166 28, Czech Republic;}
 {The Czech Academy of Sciences, Institute of Chemical Process Fundamentals,  Department of Molecular Modelling, 165 02 Prague, Czech Republic}}

\begin{abstract}
We propose a simple scaling theory describing critical effects at rounded meniscus osculation transitions which occur when the Laplace radius of a
condensed macroscopic drop of liquid coincides with the local radius of curvature $R_w$ in a confining parabolic geometry. We argue that the exponent
$\beta_{\rm osc}$ characterising the scale of the interfacial height $\ell_0 \propto R_w^{\beta_{\rm osc}}$ at osculation, for large $R_w$, falls
into two regimes representing fluctuation-dominated and mean-field like behaviour, respectively. These two regimes are separated by an upper critical
dimension, which is determined here explicitly and which depends on the range of the intermolecular forces. In the fluctuation-dominated regime,
representing the universality class of systems with short-ranged forces, the exponent is related to the value of the interfacial wandering exponent
$\zeta$ by $\beta_{\rm osc}=3\zeta/(4-\zeta)$. In contrast, in the mean-field regime,  which has not been previously identified, and which occurs for
systems with longer ranged forces (and higher dimensions), the exponent $\beta_{\rm osc}$ takes the same value as the exponent $\beta_s^{\rm co}$ for
complete wetting which is determined directly by the intermolecular forces. The prediction $\beta_{\rm osc}=3/7$ in $d=2$ for systems with
short-ranged forces (corresponding to $\zeta=1/2$) is confirmed using an interfacial Hamiltonian model which determines the exact scaling form for
the decay of the interfacial height probability distribution function. A numerical study in $d=3$, based on a microscopic model Density Functional
Theory, determines that $\beta_{\rm osc} \approx \beta_s^{\rm co}\approx 0.326$ close to the predicted value $1/3$ appropriate to the mean-field
regime for dispersion forces.
\end{abstract}

\maketitle

\section{Introduction}

It has long been recognised that fluids adsorbed at solid substrates display a wealth of new physical phenomena that are not present in the bulk.
These include wetting and prewetting transitions at planar walls \cite{schick, dietrich, sullivan} and capillary condensation or evaporation for
confinement in pores and between parallel plates \cite{evans90}, which have received extensive theoretical and experimental attention. By sculpting
the solid surface, which is now possible in the laboratory,  many more examples of surface phase transitions can be induced even in rather simple
geometries. For example, wedge filling is an example of an interfacial phase transition that is distinct from wetting \cite{hauge, rejmer, wood99,
abraham02, delfino, binder03, bernardino, our_prl, our_wedge}. Also, by merely capping a capillary the ensuing condensation can be changed from first
order to continuous  \cite{darbellay,evans_cc,tasin,mistura,hofmann, schoen,mal_groove,parry_groove,mistura13,our_groove,monson,fan,het_groove,
bruschi2, fin_groove_prl}. As well as being of interest to the fundamental statistical mechanical theory of inhomogeneous fluids and surface phase
transitions, these studies are also of relevance to microfluidics, for example.

A particularly simple example of a sculpted surface is one which is completely wet (corresponding to zero contact angle) and contoured to the shape
of a paraboloid or parabolic groove. Previous theoretical \cite{nature, carlos} and experimental \cite{exp1, exp2} studies of adsorption isotherms on
this substrate have focused on the geometry dominated growth which occurs as the bulk pressure is increased towards saturation. However, in a recent
paper \cite{osc} we pointed out that an additional rounded phase transition -- which we termed meniscus osculation -- occurs when the pressure is
tuned so that the radius of curvature of the meniscus coincides with the geometrical radius of curvature of the parabola. This marks the value of the
pressure at which the adsorption changes from being microscopic, determined by intermolecular forces or interfacial fluctuations, to being
macroscopic due to the local condensation of a liquid drop. Meniscus osculation offers another example of fluid interfacial behaviour showing
non-trivial scaling and critical effects which is related to but distinct from wetting, filling and capillary condensation.

In this paper we develop a comprehensive scaling theory for critical effects occurring at meniscus osculation and, in particular, determine the value
of the upper critical dimension which distinguishes a mean-field regime from a fluctuation-dominated one. The scaling properties which characterise
the adsorption are very different in these two regimes and are related to the underlying wetting properties via distinct critical exponent
identities. This improves upon our earlier analysis which did not identify the upper critical dimension or the mean-field regime. Two explicit
calculations, one mesoscopic and the other microscopic, are presented which determine the value of the osculation critical exponent and verify that
there are indeed two separate fluctuation regimes. More specifically, we show that analogous to the theory of complete wetting \cite{lipowsky84,
lipowsky85, fisher}, meniscus osculation shows two scaling regimes; one, which is fluctuation-dominated, characterised by universal critical
exponents that are related to the value of the wandering exponent $\zeta$, which characterises the scaling relation
\begin{equation}
 \xi_\perp \propto \xi_\parallel^\zeta\,,
 \end{equation}
between the perpendicular and parallel correlation length for planar interfaces \cite{fisher}. There is also a mean-field regime where the exponents
are sensitive to the range of the intermolecular forces where fluctuation effects are negligible. The values of the critical exponents in the
fluctuation-dominated regime, its dependence on $\zeta$, and also the value of the upper critical dimension, are different to that of the complete
wetting transition. A central result of our paper is that the value of the upper critical dimension for meniscus osculation is given by
\begin{equation}
 d^*_{\rm osc}=3-\frac{8}{3r+4}\,,
\end{equation}
where $r=2$ corresponds to dispersion forces and $r=\infty$ corresponds to short range forces. Our predictions are supported in two dimensions
($d=2$) using a droplet model treatment of an effective interfacial Hamiltonian \cite{abraham}. This determines explicitly the tail of the
probability distribution for the interfacial height above the groove bottom and identifies that the osculation critical exponent takes the value
$\beta_{\rm osc}=3/7$ in the fluctuation-dominated regime -- confirming an earlier scaling predictions, which is understood to be valid only in the
fluctuation-dominated regime \cite{osc}. In three dimensions ($d=3$) numerical studies based on a microscopic Density Functional Theory (DFT) with
dispersion forces determines that $\beta_{\rm osc} \approx 0.326$ which is close to the expectation of our scaling theory, $\beta_{\rm
osc}=\beta_s^{\rm co}=1/3$, within the mean-field regime.

Our paper is arranged as follows. We begin with a recap of the scaling theory of the fluctuation regimes and the critical singularities for complete
wetting transitions at planar walls before developing a crossover scaling theory which identifies the relevant length scales and critical
singularities at meniscus osculation. A general scaling theory is presented which, similar to complete wetting, separates critical singularities into
fluctuation-dominated and mean-field regimes. Explicit examples which confirm these predictions in $d=2$ and $d=3$ are presented. We finish our paper
with a brief summary and discussion of possible further work.

\section{Scaling theory for complete wetting}

\begin{figure}
 \includegraphics[width=8cm]{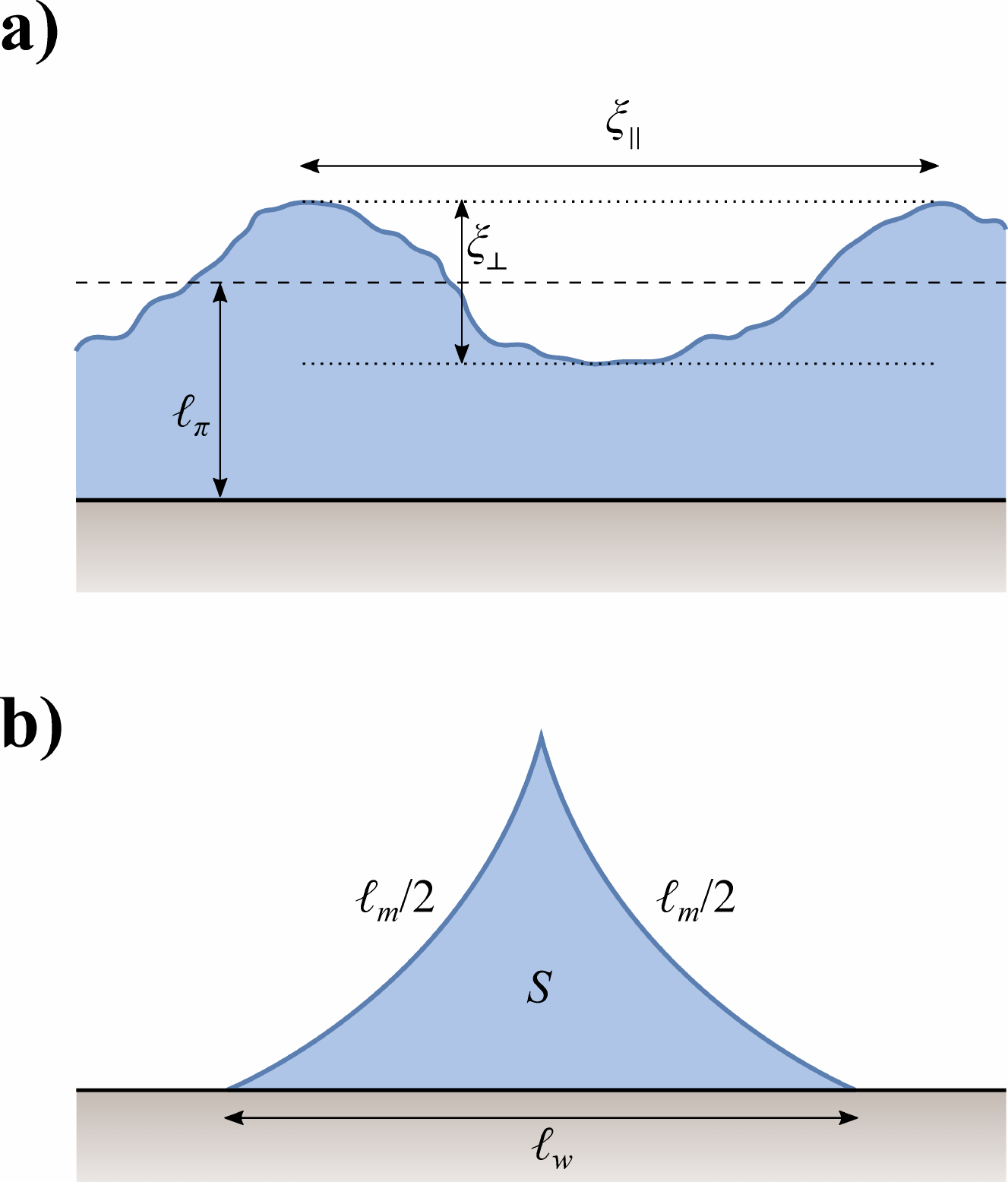}
\caption{ a) Schematic illustration of the equilibrium interfacial thickness $\ell_\pi\propto \delta p^{-\beta_s^{\rm co}}$, parallel correlation
length $\xi_\parallel\propto \delta \mu^{-\nu_\parallel^{\rm co}}$ and interfacial roughness $\xi_\perp \propto \xi_\parallel^\zeta$ for complete
wetting by liquid (blue) at a planar wall-gas interface. b) Illustration of a droplet configuration in $d=2$ constrained to pass through a point at
height $\ell\gg\ell_\pi$, i.e. one a scale much larger than the length scales shown in Fig.~1a), far above the wall. The free-energy cost $\Delta
F(\ell)\propto \delta p^{1/2} \ell^{3/2}$ of the droplet determines the asymptotic scaling form of the interfacial height probability distribution
function $P(\ell)$, identifying explicitly that for systems with short-ranged forces $\beta_s^{\rm co}=1/3$. The droplet area $S$, interfacial length
$\ell_m$ and the length of contact with the wall $\ell_w$ are shown. } \label{fig1}
\end{figure}

To begin, we recall some details of the well developed fluctuation theory of complete wetting \cite{lipowsky84, lipowsky85, fisher, schick} which we
will need in our analysis of meniscus osculation. The complete wetting transition refers to the divergence in the adsorption, $\Gamma$, of liquid at
a planar wall-gas interface (say), as the pressure $p$ (or chemical potential $\mu$) is increased to saturation $p_{\rm sat}$, above a wetting
temperature, i.e. when the macroscopic contact angle $\theta=0$. As $\delta p=p_{\rm sat}-p \to 0$, a number of length scales diverge, in particular
 \begin{equation}
 \ell_\pi \propto \delta
p^{-\beta_s^{\rm co}}\,; \hspace {1cm}\xi_\parallel \propto \delta p^{-\nu_\parallel^{\rm co}}\,. \label{ell_pi}
\end{equation}
Here, $\ell_\pi$ is the wetting layer thickness which is related to the adsorption, $\Gamma=\Delta\rho\ell_\pi$, where $\Delta\rho$ is the difference
between bulk liquid and gas densities, and $\xi_\parallel$ is the parallel correlation length arising from the build-up of capillary-wave-like
fluctuations near the unbinding liquid-gas interface which leads also to the divergence of the interfacial roughness $\xi_\perp$ -- see
Fig.~\ref{fig1}a. For pure systems, as pertinent to wall-fluid interfaces, it is well established that the wandering exponent $\zeta=(3-d)/2$ for
dimension $d<3$, (with $\xi_\perp^2 \propto \ln \xi_\parallel$ in $d=3$ corresponding to $\zeta=0$) with its value also known for impure systems
(most commonly, random-bond and random-field disorder) \cite{fisher}. For complete wetting, an exact sum-rule determines that
$\partial\Gamma/\partial\mu \propto \xi_\parallel^2$, leading to the exact exponent relation \cite{row, hend, evans}
\begin{equation}
1+\beta_s^{\rm co}=2\nu_\parallel^{\rm co} \label{exp_id}\,.
\end{equation}
 The values of the critical exponents can be determined, quite generally, from analysis of the simple interfacial model \cite{lipowsky84, lipowsky85}
\begin{equation}
H[\ell]= \int d{\bf x} \left[\frac{\gamma}{2} (\nabla\ell)^2+W(\ell)\right]\,, \label{ham}
 \end{equation}
where $\ell({\bf x})$ is the interfacial coordinate (measuring the local height of the wetting layer at the position ${\bf x}$ along the wall),
$\gamma$ is the surface tension which resists interfacial fluctuations and
\begin{equation}
W(\ell)=\delta p \ell+\frac{A}{\ell^r}+\cdots \label{bind}
\end{equation}
is the binding potential which includes the effect of intermolecular forces characterised by the exponent $r$ (with $A$ a Hamaker constant), which
maybe derived from more microscopic theory \cite{dietrich}. Heuristic scaling arguments suggest that the interfacial wandering leads to an effective
entropic repulsion, decaying as $\ell^{-\tau}$ where $\tau=2(1-\zeta)/\zeta$, which competes with the direct intermolecular contribution in
$W(\ell)$, leading to two scaling regimes:

{\it Fluctuation-dominated regime}. For $r>\tau$, fluctuations dominate leading to scaling behaviour, characterised by  $\ell_\pi \propto \xi_\perp
\propto \xi_\parallel^\zeta$, with universal non-classical critical exponents \cite{lipowsky84}
 \begin{equation}
 \beta_s^{\rm co}=\frac{\zeta}{2-\zeta}\,, \hspace{1cm} \nu_\parallel^{\rm co}=\frac{\zeta}{2-\zeta}\,. \label{beta_s}
 \end{equation}
The dependence on the wandering exponent $\zeta$ here is quite general and applies also to impure systems. Thus in $d=2$ the critical exponent
$\beta_s^{\rm co}=1/3$ for pure systems ($\zeta=1/2$), while $\beta_s^{\rm co}=1/2$ for systems with random-bond disorder ($\zeta=2/3$).

{\it Mean-field regime}. For $r<\tau$, on the other hand, the intermolecular forces dominate leading to mean-field-like critical behaviour for which
$\ell_\pi\gg\xi_\perp$ with critical exponents
 \bb
\beta_s^{\rm co}=\frac{1}{1+r}\,, \hspace{1cm} \nu_\parallel^{\rm co}=\frac{2+r}{2(1+r)}\,,
 \ee
which follow from simple minimization of the binding potential.

For fixed $r$, and systems with just thermal disorder, these regimes determine the upper critical dimension
\begin{equation}
 d^*=3-\frac{4}{r+2}\,,
\end{equation}
below which fluctuations dominate and above which they are negligible \cite{lipowsky84}. In $d=3$ and with dispersion forces ($r=2$) this implies
$\beta_s^{\rm co}=1/3$ (and $\nu_\parallel^{\rm co}=2/3$), as predicted many years ago by Derjaguin and which has been confirmed exhaustively in
numerous experiments \cite{sullivan, schick}.

\begin{figure}
 \includegraphics[width=8cm]{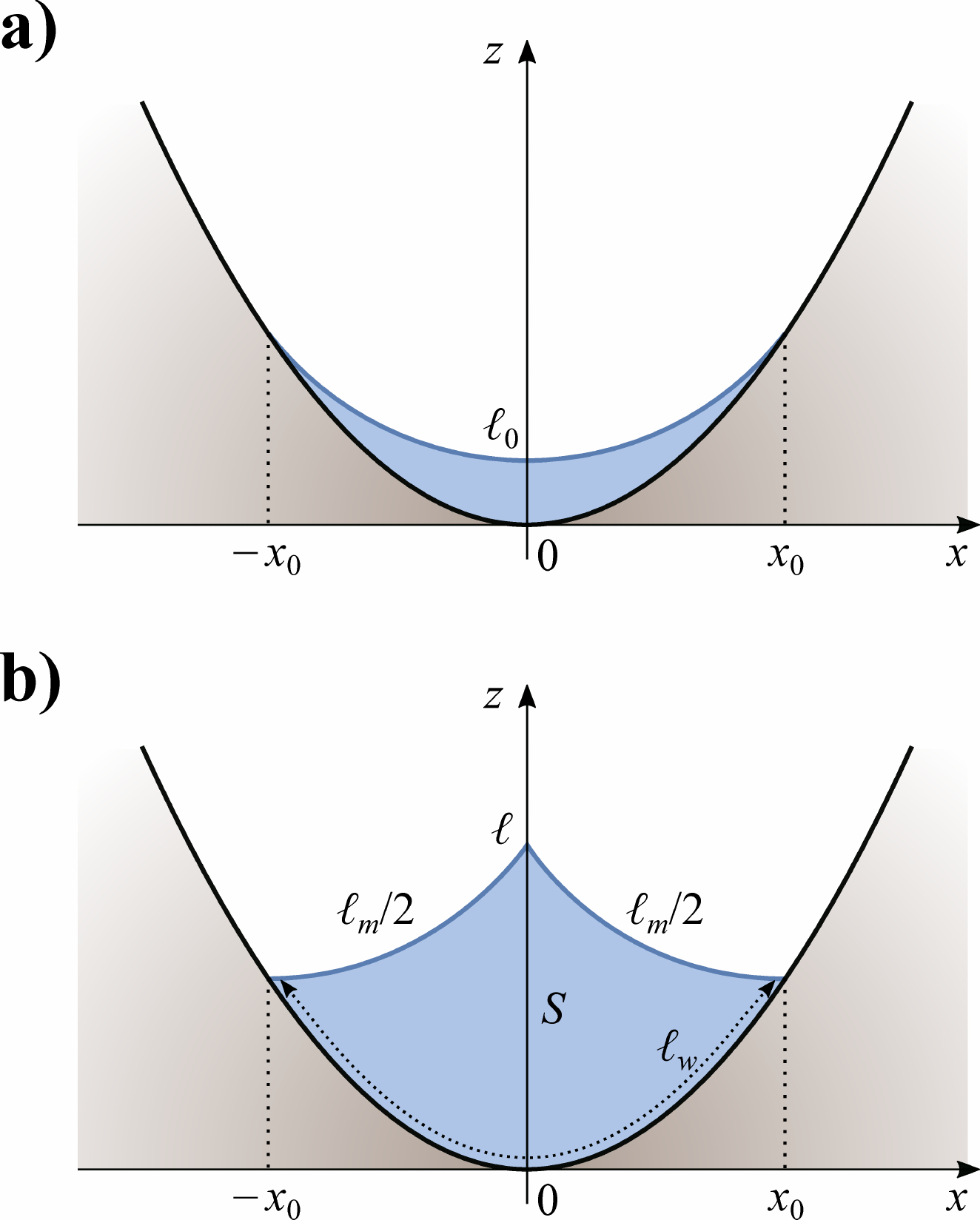}
\caption{a) Schematic illustration of the height $\ell_0=(R-R_w)^2/2R_w$ and width $x_0=\sqrt{R^2-R_w^2}$ of a macroscopic drop adsorbed in a
parabolic groove in the regime $R>R_w$ close to bulk coexistence (with $R=\gamma/\delta p$ the Laplace radius). For $R<R_w$ there is no macroscopic
drop and the adsorption remains microscopic arising from interfacial fluctuations or intermolecular forces. b) Illustration of the constrained
fluctuation droplet configuration in $d=2$ which determines the free energy and asymptotic scaling form of the PDF for the interfacial height (above
the groove bottom) allowing us to identify that $\beta_{\rm osc}=3/7$. } \label{fig2}
\end{figure}

In $d=2$ these heuristic expectations are also fully confirmed using discrete and continuum interfacial Hamiltonians  \cite{lipowsky85, abraham86}.
The partition function for the interfacial model (\ref{ham}) can be determined exactly using continuum transfer matrix methods, equivalent to solving
the eigenfunctions and eigenvalues of the Schrodinger-like equation
\begin{equation}
-\frac{1}{2\beta^2\gamma}\psi_n''(\ell)+W(\ell)\psi_n(\ell)=E_n \psi_n(\ell)\,,
\end{equation}
where $\beta=1/k_BT$ which we hereafter set to unity. This elegant transfer matrix method determines, for example, that the probability distribution
function (PDF) of finding the interface at height $\ell$ is $P(\ell)=|\psi_0(\ell)|^2$, which determines both $\ell_\pi$ and $\xi_\perp$, and also
identifies that $\xi_\parallel=1/(E_1-E_0)$. For the systems with short-ranged forces (representing the scaling regime for $r>2$), the eigenfunctions
are given by $\psi_n(\ell) \propto {\rm Ai}((2\gamma\delta p)^{1/3}\ell+\lambda_n)$, where the $\lambda_n$ are the zeros of the Airy function ${\rm
Ai}(x)$ with corresponding eigenvalues $E_n=2^{-1/3}|\lambda_n|(\delta p)^{2/3}$. The power-law dependence on $\delta p$ within $P(\ell)$ immediately
determines that $\beta_s^{\rm co}=1/3$, consistent with (\ref{beta_s}) on setting $\zeta=1/2$ as appropriate to $d=2$. The decay of the Airy function
then determines that far from the wall the PDF decays as
 \bb
P(\ell)\propto \frac{1}{\ell^{1/2}}{\rm exp}\left[-\frac{4}{3}(2\gamma\delta p)^{1/2}\ell^{3/2}\right]\,, \label{tm}
 \ee
which, of course, still reveals the value of the complete wetting exponent $\beta_s^{\rm co}$. This asymptotic behaviour is completely consistent
with the \emph{droplet} model expectation that the PDF decays as \cite{abraham}
\begin{equation}
P(\ell)\propto e^{-\Delta F(\ell)}\,,
\end{equation}
where $\Delta F(\ell)$ is the free-energy cost, in units of $k_B T$, of forming a constrained droplet of liquid which rises above the wall forming a
cusp at height $\ell$ (see Fig.~\ref{fig1}b). For large $\ell$ this free-energy cost is macroscopic and is simply given by
\begin{equation}
\Delta F(\ell)=\delta pS+\gamma(\ell_m-\ell_w)\,,
\end{equation}
where $S$ is the area of the droplet, $\ell_m$ is the interfacial length and $\ell_w$ the length of contact with the wall (see Fig.~\ref{fig1}b).
Either side of the cusp the droplet has a parabolic shape with curvature $\delta p/\gamma$ and a very simple calculation shows that both the area and
interfacial length contributions to the free-energy cost are the same determining that $\Delta F(\ell)=\frac{4}{3}(2\gamma\delta p)^{1/2}\ell^{3/2}$,
in precise agreement with the transfer-matrix analysis (\ref{tm}). The algebraic pre-factor in (\ref{tm}) is related to the interfacial wandering at
the points of contact of the droplet with the wall, similar to discussions of the magnetization profile in the semi-infinite Ising model
\cite{abraham}. We shall use this droplet model trick later in application to the meniscus osculation transition.

\section{Scaling theory for meniscus osculation}

We now turn our attention to the adsorption of fluid near a completely wet wall which has the shape of a parabolic groove (or parabolic pit) of
cross-section
 \bb
Z(x)=\frac{x^2}{2R_w}\,, \label{wall}
 \ee
where $R_w$ is the geometrical radius of curvature at the bottom. The adsorption falls into two regimes depending on the deviation from bulk
coexistence $\delta p$. Close to coexistence, when $R>R_w$ where $R =\gamma/\delta p$ is the Laplace radius, the groove induces the local
condensation of a macroscopic liquid drop near the bottom. The drop is characterised by a circular meniscus of radius $R$ that meets the walls
tangentially (since $\theta =0$) -- see Fig.~\ref{fig2}a. The size of this drop is determined trivially. For example, the local height $\ell_0$ and
lateral extension $x_0$ of the drop are given by \cite{osc}
 \bb
\ell_0=\frac{(R-R_w)^2}{2R_w}\,,  \label{l0}
 \ee
 and
 \bb
 x_0=\sqrt{R^2-R_w^2}\,, \label{x0}
 \ee
respectively. As we approach coexistence these diverge as $\ell_0\propto R^2$  and $x_0\propto R$ which is the expected geometry-dependent behaviour
for the adsorption in a parabola \cite{nature}. However, these results also indicate that these length scales vanish as the pressure is reduced and
$R\to R_w$ and hence that further away from coexistence, corresponding to pressures such that $R<R_w$, there is no local condensation and the
adsorption of fluid is microscopic. We refer to the vanishing of the meniscus at $R=R_w$ as meniscus osculation. At a macroscopic level this is a
continuous surface phase transition associated with a singular contribution to the surface free-energy which vanishes as $F_{\rm sing}\approx (R-
R_w)^{7/2}$ \cite{osc}.

Beyond macroscopic considerations meniscus osculation must correspond to a rounded phase transition since there must still be some residual
microscopic adsorption in the pressure regime $\delta p>\gamma/R_w$. The rounding at meniscus osculation leads to novel scaling behaviour
characterising the influence of the geometry on the fluid adsorption at the borderline of the macroscopic and microscopic regimes. Consider, for
example, the height, $\ell_0$, of the liquid interface above the groove bottom exactly at osculation $R=R_w$. Since the wall is completely wet,
$\ell_0$ must increase with $R_w$ (maintaining the condition that $R_w=R$) allowing us to define  an osculation exponent $\beta_{\rm osc}$:
 \bb
 \ell_0\propto R_w^{\beta_{\rm osc}}\,, \label{beta_osc}
 \ee
which characterises the local divergence of the film thickness as we flatten the groove and recover the infinite adsorption of a wet planar wall.

To determine the value of this exponent we suppose that the macroscopic osculation transition is rounded over a microscopic scale $\lambda\ll R$ to
be determined. It is natural to speculate that this must be related to length scales which characterise the underlying complete wetting phenomena
discussed above. Crossover scaling then suggests that, in the vicinity of the phase boundary $R\approx R_w$, the macroscopic results (\ref{l0}) and
(\ref{x0}) are modified as
 \bb
 \ell_0=\frac{(R-R_w)^2}{2Rw}{\cal{L}}_{\rm osc}\left(\frac{R-R_w}{\lambda_{\rm osc}}\right)\label{l0_scale}
 \ee
 and
 \bb
 x_0=\sqrt{R^2-R_w^2}{\cal{X}}_{\rm osc}\left(\frac{R-R_w}{\lambda_{\rm osc}}\right)\,, \label{x0_scale}
 \ee
where ${\cal{L}}_{\rm osc}(x)$ and ${\cal{X}}_{\rm osc}(x)$ are scaling functions of the dimensionless variable $x=(R-R_w)/\lambda$. Note that the
microscopic length scale $\lambda$ is still allowed to diverge as bulk coexistence is approached but we require that is always much smaller than the
purely macroscopic length-scale $R$. We require that both scaling functions tend to unity as $x \to \infty$ and that both vanish as $x\to -\infty$ in
order to recover the macroscopic results. The crossover length scale determines the values of $\ell_0$ and $x_0$ at the macroscopic phase boundary
$R=R_w$. In order that these are finite and non-vanishing we require that ${\cal{L}}_{\rm osc}\sim |x|^{-2}$ and ${\cal{X}}_{\rm osc}\sim |x|^{-1/2}
$ as $x\to0$, which identifies that
\begin{equation}
\ell_0 \propto \frac{\lambda^2}{R}\,,\hspace{1cm} x_0 \propto \sqrt{R \lambda}\,;\hspace{1cm} R=R_w\,.  \label{scaling}
\end{equation}
From these we can immediately rule out that $\lambda$ is similar to the planar wetting layer thickness since in that case $\ell_0$ does not diverge
with $R_w$ as required. In Ref.~\cite{osc} we argued there were likely two possibilities. The simplest, and perhaps most natural, hypothesis is that
$\lambda\sim \xi_\parallel$. This is indeed the length scale which controls the crossover scaling and rounding at meniscus depinning transitions
\cite{md1, md2} and also wetting on rough surfaces (where it is sometimes referred to as the healing length \cite{andelman}). With this ansatz it
follows from (\ref{exp_id}) and (\ref{scaling}) that $\ell_0\propto \ell_\pi$ so that $\beta_{\rm osc}=\beta_s^{\rm co}$, i.e. the parabola doesn't
significantly enhance the film thickness compared to that at a planar wall, although it is likely to be a multiple of it. However, there is an
alternate possibility that is also justifiable, which is that deep in the pre-osculation regime ($R\ll R_w$) the influence of the geometry on the
film thickness is to shift and reduce the effective pressure from $\delta p$ to $\delta p -\gamma/R_w$. This geometrically induced shift would be
consistent with the effective increase in the pressure, which is known for wetting on the outside of a sphere or cylinder \cite{lipowsky87, bieker,
stewart, morgan, nold}. This means that as $R_w \to \infty$ the local height tends to $\ell_0 \propto (1/R-1/R_w)^{-\beta_s^{\rm co}}$, which is only
compatible with the scaling hypothesis (\ref{l0_scale}) if $\lambda^{2+\beta_s^{\rm co}}\propto R_w^{1+2\beta_s^{\rm co}}R^{\beta_s^{\rm co}}$. With
this identification for the rounding length scale it follows from (\ref{scaling}) that the value of $\ell_0$ is much larger than $\ell_\pi$ and
diverges on approaching coexistence with exponent $\beta_{\rm osc}=3\beta_s^{\rm co}/(2+\beta_s^{\rm co})$, which is larger than $\beta_s^{\rm co}$.

Here, we argue that both these possibilities are realised and that they are characteristic of the rounding occurring in two different scaling regimes
demarcated by an upper critical dimension. Consider, for example, the rounding and scaling resulting from the assertion that the substrate curvature
decreases the effective pressure $\delta p$ to $\delta p-\gamma/R_w$. It is natural to assume that this purely geometrical consideration occurs for
systems with sufficiently short-ranged forces where the influence of intermolecular forces can be neglected. This is somewhat analogous to the
``wedge covariance" known for wetting and filling phenomena in systems with short-ranged forces (in both pure and impure systems) which exactly
relates thermodynamic observables in a wedge (with tilt angle $\alpha$) to that at a planar wall via an effective shift in the contact angle $\theta
\to \theta-\alpha$ \cite{parry02}. Combining the exponent relation $\beta_{\rm osc}=3\beta_s^{\rm co}/(2+\beta_s^{\rm co})$ with the result
$\beta_s^{\rm co}=\zeta/(2-\zeta)$ for short-ranged complete wetting leads to the explicit identification $\beta_{\rm osc}=3\zeta/(4-\zeta)$ for
meniscus osculation. This is greater than the corresponding value $\beta_s^{\rm co}=\zeta/(2-\zeta)$ for complete wetting for all $\zeta<1$ -- that
is, for all dimensions above the lower critical dimension for bulk phase separation. Nevertheless, this is precisely what we should expect if we
assume that the phenomena arise from interfacial fluctuations, since in that case we can also anticipate that $\ell_0 \propto x_0^\zeta$ -- that is
the wandering exponent is \emph{unchanged} by the geometry. Combining this expectation with the crossover scaling result  (\ref{scaling}) identifies
that, at osculation, $\lambda \propto R^{(2+\zeta)/(4-\zeta)}$, which consistently and \emph{independently} identifies that $\beta_{\rm
osc}=3\zeta/(4-\zeta)$. This, we conjecture, is the appropriate rounding length-scale and value of the osculation exponent for systems with
sufficiently short-ranged forces. However, this scaling cannot apply universally. As we increase the dimensionality the value of $\zeta$ decreases
and eventually the osculation critical exponent reaches the value $\beta_{\rm osc}=1/(r+1)$ implying that $\ell_0\propto\ell_\pi$ and
$\lambda\propto\xi_\parallel$. Since $\beta_{\rm osc}$ cannot take a smaller value than the corresponding value of $\beta_s^{\rm co}$ (the confining
geometry cannot diminish the adsorption) it is natural to assume that this scaling applies also in all higher dimensions. Thus, analogous to complete
wetting we conjecture that meniscus osculation falls into one of two scaling regimes:

{\it Fluctuation-dominated regime}.  For $r>4(1-\zeta)/3\zeta$, fluctuations dominate and the osculation exponent takes the universal value
 \bb
 \beta_{\rm osc}=\frac{3\zeta}{4-\zeta}\,.
 \ee
In this scaling regime $\ell_0\sim x_0^\zeta$,  implying that the geometry significantly enhances the adsorption, such that $\ell_0\gg\ell_\pi$. The
crossover scaling and rounding of the meniscus osculation transition is controlled by a length scale $\lambda\approx R^{(2+\zeta)/(4-\zeta)}$, which
is larger than the corresponding value of $\xi_\parallel$ (at this pressure). Thus in $d=2$ we predict that the meniscus osculation is characterized
by the exponent $\beta_{\rm osc}=3/7$ for pure systems ($\zeta=1/2$) and $\beta_{\rm osc}=3/5$ for random-bond disorder ($\zeta=2/3$). These contrast
with the corresponding prediction for compete wetting $\beta_s^{\rm co}=1/3$ (for $\zeta=1/2$) and  $\beta_s^{\rm co}=1/2$ (for $\zeta=2/3$).

{\it Mean-field regime}. For $r<4(1-\zeta)/3\zeta$, the intermolecular forces dominate and the osculation exponent takes the value
 \bb
 \beta_{\rm osc}=\frac{1}{r+1}\,,
 \ee
which is identical to the value of $\beta_s^{\rm co}$. This implies that the local interfacial height  scales with the wetting layer thickness, i.e.
$\ell_0\propto\ell_\pi$. We anticipate that in general the constant of proportionality is greater than unity, so that the geometry still enhances the
local adsorption of fluid. The rounding of the phase transition in this regime is controlled by a crossover length scale
$\lambda\propto\xi_\parallel$.

For fixed value of $r$, these two scaling regimes  identify that for pure systems that the upper critical dimension is
\begin{equation}
 d^*_{\rm osc}=3-\frac{8}{3r+4}\,,
\end{equation}
which is larger than the upper critical dimension for complete wetting, except for systems with purely short-ranged forces, for which $d^*=d^*_{\rm
osc}=3$.

\section{Model calculations}

To finish our article, we test these predictions for the two cases that are most relevant to experiments and studies of microscopic models: $d=2$
with short-ranged forces ($r=\infty$) and $d=3$ with dispersion forces ($r=2$). In both these cases the values of the complete wetting exponents are
identical with $\beta_s^{\rm co}=1/3$ and $\nu_\parallel^{\rm co}=2/3$ -- although these correspond to distinct fluctuation and mean-field regimes
respectively. The predictions of the scaling theory developed above are that in $d=2$ the osculation exponent $\beta_{\rm osc}=3/7$, different to
that for complete wetting, while in $d=3$ it remains $\beta_{\rm osc}=\beta_s^{\rm co}=1/3$.

\subsection{$d=2$, short-ranged forces}

In $d=2$ we may study meniscus osculation using a continuum  interfacial Hamiltonian adopting the same droplet model method described earlier for
complete wetting. This, we anticipate, will exactly determine the scaling form of the asymptotic probability distribution for the local interfacial
height above the groove bottom. That is, we assume that
 \bb
P_{\rm osc}(\ell) \propto {\rm exp}[-\Delta F_{\rm osc}(\ell)]\,, \label{posc}
 \ee
where $\Delta F_{\rm osc}(\ell)$ is the free-energy cost (in units of $k_BT$) for an interfacial fluctuation that forms a droplet which is
constrained to pass through a point at height $\ell$ at $x=0$ -- see Fig.~\ref{fig2}b. Since no direct intermolecular forces are present the
free-energy cost of this droplet fluctuation is again given by
\begin{equation}
\Delta F_{\rm osc}(\ell)=\delta p \mathcal{S}+ \gamma(\ell_m-\ell_w)\,,
\end{equation}
where $\mathcal{S}$ is the area, $\ell_m$ is the interfacial length and $\ell_w$ the length of contact with the parabolic wall. The droplet has the
shape of a symmetric cusp formed from two circular menisci of Laplace radius $R$, centered at $x=\pm \xi$ that meet the walls tangentially at $x=\pm
x_0$. For $x>0$ the local interfacial height is therefore described by the function
\begin{equation}
\ell(x)=\ell -\sqrt{R^2-(x-\xi)^2}+ \sqrt{R^2-\xi^2}\,,
\end{equation}
which we may expand keeping terms of quartic order
\begin{equation}
\ell(x)=\ell - \frac{\xi^2}{2R} -\frac{\xi^4}{8R^3}+\frac{(x-\xi)^2}{2R}+\frac{(x-\xi)^4}{8R^3}\,,
\end{equation}
which is the order required to determine the scaling behaviour.
 We now sit exactly at osculation $R=R_w$ and
define reduced variables $\tilde \ell=\ell/R$, $\tilde \xi=\xi/R$ and $\tilde x_0=x_0/R$. Matching the interface and wall heights,
$\ell(x_0)=Z(x_0)$, and derivatives $\ell'(x_0)=Z'(x_0)$ determines that
\begin{equation}
\tilde x_0=\tilde \xi+(2\tilde \xi)^{1/3}
\end{equation}
and
\begin{equation}
\tilde \ell=\frac{3}{8 } \tilde \xi^{4/3}+\tilde \xi^2+\tilde \xi^4\,.
\end{equation}
Using these it is a straightforward matter to determine the interfacial area $\mathcal{S}=2\int_0^{x_0}[(\ell(x)-\Psi(x)]dx$, yielding
\begin{equation}
 \frac{\mathcal{S}}{R^2}=\frac{3}{10}(2 \tilde \xi)^{5/3}+\frac{3}{16}(2 \tilde\xi)^{7/3}+\cdots\,,
 \end{equation}
 where the higher-order terms are of
${\mathcal{O}}(\tilde\xi^3)$ which may be neglected. Similarly, the surface terms, relating to the excess length of the droplet,
$\ell_m-\ell_w=2\int_0^{x_0} dx(\sqrt{1+\ell'(x)^2}-\sqrt{1+\psi'(x)^2})$ follow as
\begin{equation}
\frac{\ell_m-\ell_w}{R}=-\frac{3}{10}(2 \tilde \xi)^{5/3}-\frac{1}{8}(2 \tilde \xi)^{7/3}+\cdots\,,
\end{equation}
where the higher-order terms are also $\mathcal{O}(\tilde \xi^{3})$. The leading-order terms in the area and length contributions cancel implying
that, exactly at osculation, the free-energy cost of the drop scales with the local height as
 \begin{equation}
 \Delta F_{\rm osc}=\frac{\gamma}{4}  R^{-\frac{3}{4}}\left(\frac{8\ell}{3}\right)^{\frac{7}{4}}\,.
 \end{equation}
Substitution to (\ref{posc}) then immediately determines that the osculation exponent takes the predicted value
 \begin{equation}
 \beta_{\rm osc}= \frac{3}{7}\,.
 \end{equation}
Fluctuations are important at this rounded phase transition so that, for example, the interfacial roughness also scales as $\xi_\perp \propto
R_w^{3/7}$. We anticipate that in Eq.~(\ref{posc}) there is also an algebraic pre-factor  associated with the interfacial wandering of the points of
contact, similar to the droplet model for complete wetting, although this is not relevant to the scaling behaviour and the identification of
$\beta_{\rm osc}$.

\subsection{$d=3$, long-ranged forces}

\begin{figure}
\includegraphics[width=8cm]{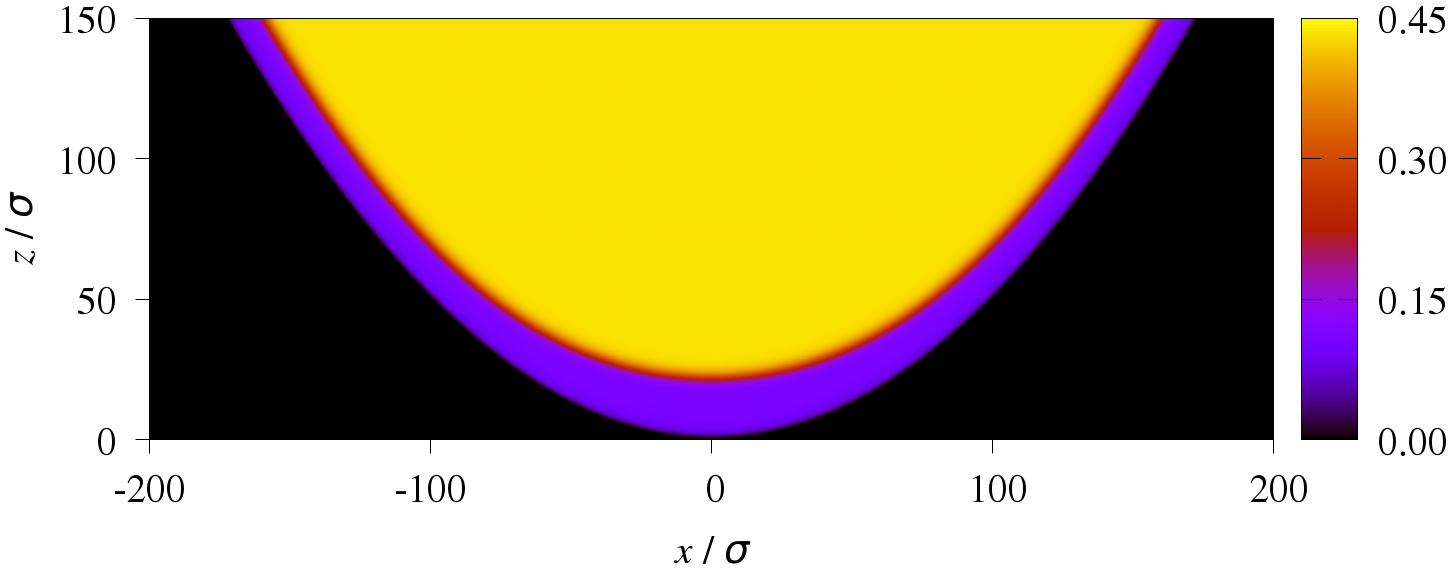}
\caption{Numerical DFT results for the equilibrium density profile $\rhor$ at osculation  for a completely dry parabolic wall ($R_w=100\,\sigma$) in
contact with a bulk liquid showing the preferential adsorption of low density gas at the bottom. } \label{fig3}
\end{figure}

\begin{figure}
\includegraphics[width=8cm]{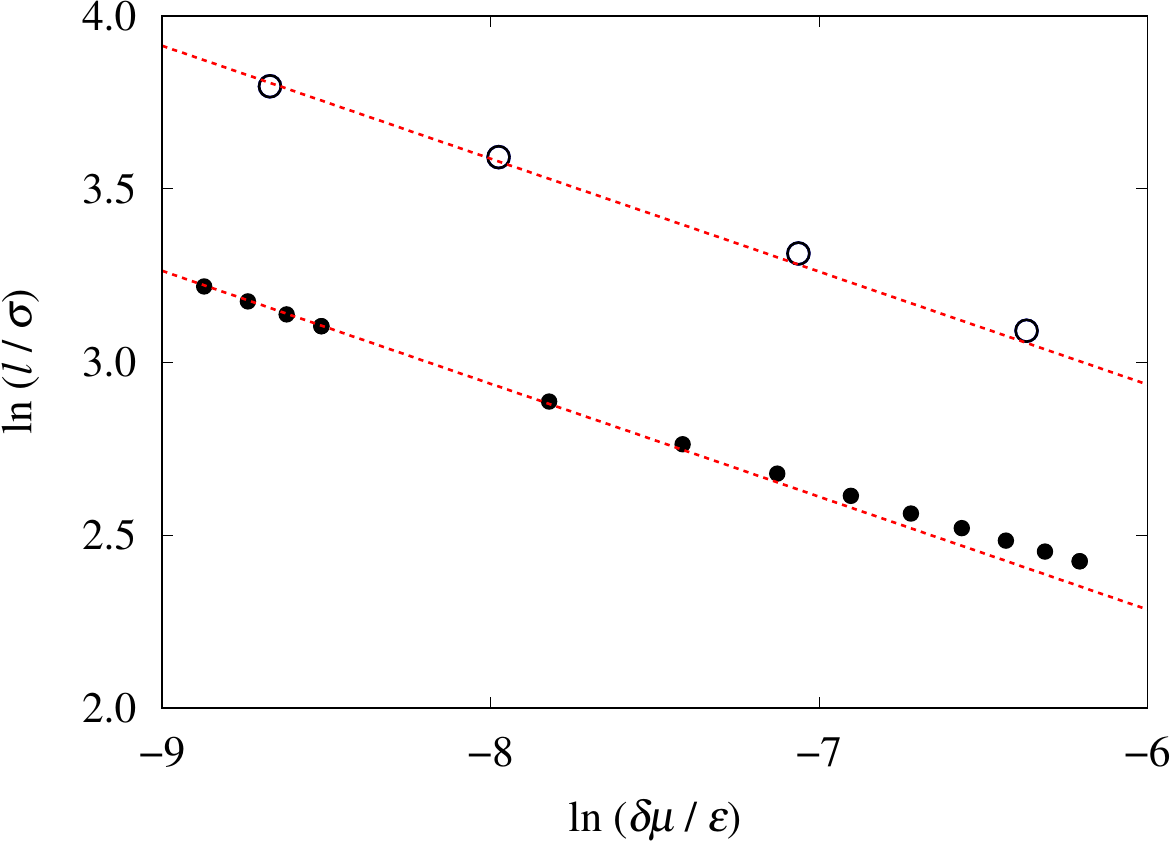}
\caption{Log-log plot showing the growth of the interfacial height $\ell_0$ (open circles) )and the planar wetting thickness $\ell_\pi$ (dark
circles) for different radii of curvature $R_w$ maintaining the condition of meniscus osculation $\mu_{\rm osc}=\mu_{\rm sat}+\gamma/\Delta \rho
R_w$. The two straight lines shown are near parallel identifying that $\beta_{\rm osc}\approx \beta_s^{\rm co}\approx0.326$ with corresponding
amplitude ratio $\ell_0/\ell_\pi\approx 2$. } \label{fig4}
\end{figure}

To study meniscus osculation in $d=3$ we employ a fully microscopic, classical DFT which is based on the minimization of a grand potential functional
$\Omega[\rho]$ with respect to the density distribution of the fluid particles $\rhor$ \cite{evans79}:
   \bb
  \Omega[\rho]=F[\rho]+\int\dr\rho(\rr)\left[V(\rr)-\mu\right]\,.\label{grandpot}
 \ee
Here, $F[\rho]$ is the Helmholtz free energy functional which contains all the information about the fluid interactions, while $V(\rr)$ is the
potential of the parabolic wall whose cross-section along the $x$-$z$ plane is given by Eq.~(\ref{wall}). The wall is formed of atoms which are
distributed uniformly with a density $\rho_w$ over the whole space below its surface demarcated by the curve (\ref{wall}) assuming translation
invariance along the $y$-axis. The wall atoms interact with the fluid atoms via a purely repulsive potential, $\phi=4\varepsilon (\sigma/r)^{6}$,
hence the net wall potential is
 \bb
 V(\rr)=\rho_w\int\phi(|\rr-\tilde\rr|)\dd\tilde \rr\,,
 \ee
where the integration domain  is the volume of the wall. Here, $\varepsilon$ is the strength of the potential, while $\sigma$ is molecular radius.
The repulsive tail of the wall potential models dispersion interactions which, within the mesoscopic interfacial model (\ref{ham}), generate a
binding potential decaying asymptotically according to a power-law with $r=2$. However, we note that in this context, because the intermolecular
interaction is purely repulsive, we consider the analogous drying phenomena when the repulsive wall is brought in contact with bulk liquid. An
advantage of this is that the drying layer of gas does not exhibit volume exclusion effects allowing us to access a greater range of $R_w$ values.

The fluid-fluid interaction is modelled by a (short-ranged) truncated Lennard-Jones potential (of strength $\varepsilon$) and its contribution to the
free energy functional is described by a combination of Rosenfeld's fundamental measure theory \cite{ros} (approximating the repulsive part of the
interaction) and a simple mean-field treatment of the attractive part of the interaction. More details about the construction of the approximative
$F[\rho]$ and the numerical details of minimization of $\Omega[\rho]$ can be found in Ref.~\cite{osc} where the same fluid model has been adopted.

In order to determine the exponent $\beta_{\rm osc}$, we first found the equilibrium density profiles for various parabolic walls with different
curvatures with fixed chemical potential, $\mu_{\rm osc}=\mu_{\rm sat}+\gamma/(R_w \Delta\rho)$, ensuring that we sit right at the osculation
transition (see Fig.~\ref{fig3}). From each density profile we determined the interfacial height above the groove bottom $\ell_0$ using the
mid-density rule. In Fig.~\ref{fig4} we display the log-log dependence of $\ell_0$ with $\delta\mu_{\rm osc}$ (with $\delta\mu_{\rm osc}\equiv
\mu_{\rm osc}-\mu_{\rm sat}$) comparing it also with the corresponding divergence of the planar wetting thickness $\ell_\pi$ for the same range of
chemical potentials. This shows convincingly that $\ell_0$ and $\ell_\pi$ diverge with the same critical exponent which we estimate as $\beta_{\rm
osc}\approx \beta_s^{\rm co}\approx0.326$ in excellent agreement with the predicted value of  $\beta_{\rm osc} = 1/3$. Our results indicate that the
ration $\ell_0/\ell_\pi\approx 2$ showing that at meniscus osculation within this mean-field regime the geometry increases the amplitude of the local
adsorption but not the critical exponent.

\section{Summary}

In this paper we have developed a simple scaling theory for critical effects which arise from the rounding of the meniscus osculation transition
occurring when the Laplace pressure of a condensed macroscopic drop of liquid coincides with local radius of curvature $R_w$ in a confining parabolic
geometry. We have argued that the exponent $\beta_{\rm osc}$ characterising the scale of the interfacial height $\ell_0 \propto R_w^{\beta_{\rm
osc}}$ at osculation, falls into one of two regimes representing fluctuation-dominated and mean-field like behaviour. In the fluctuation-dominated
regime, representing the universality class of systems with short-ranged forces, the exponent is related to the value of wandering exponent by
$\beta_{\rm osc}=3\zeta/(4-\zeta)$ which is different to the relation $\beta_s^{\rm co}=2/(2-\zeta)$ pertinent for complete wetting. This exponent
relation can be understood to arise in two equivalent ways -- either by assuming that when fluctuations dominate the height $\ell_0$ and lateral size
$x_0$ of the adsorbed layer scale as $\ell_0 \sim x_0^\zeta$ or by enforcing a condition on the crossover scaling function that in the pre-osculation
regime the geometry serves to lower the effective partial pressure $\delta p \to \delta p -\gamma/R_w$. These simple scaling considerations do not
apply if the forces are sufficiently long-ranged in which case the midpoint interfacial height $\ell_0 \propto \ell_\pi$ and rounding length scale
$\lambda \approx \xi_\parallel$ are more directly and simply related to wetting length-scales. Our prediction that in $d=2$ and for short-ranged
forces the meniscus osculation exponent takes the value $\beta_{\rm osc}=3/7$ is confirmed by a droplet model calculation based on an interfacial
Hamiltonian which determines the scaling form of the asymptotic decay of the PDF for the local interfacial height. Future studies could seek to
extend this to and determine, for example, the whole PDF including the short-distance expansion near the wall which we anticipate can be related to
exact sum-rules similar to studies of continuous wetting at planar walls \cite{parry91}. In $d=3$ our DFT study indicates that in the mean-field
regime with dispersion forces the ratio of the interfacial heights $\ell_0/\ell_\pi\approx 2$. It would be interesting to see if the value of this
amplitude can be understood using simple interfacial Hamiltonian models, which also allow for the presence of long-ranged forces  \cite{tas}. This
would have implications for understanding adsorption on other types of surface \cite{nature, carlos}. Finally, the adsorption of fluids in substrates
with parabolic pits has been considered experimentally previously \cite{exp1, exp2}  although the meniscus osculation was not addressed. We hope that
the present work stimulates such studies.

\begin{acknowledgments}
\noindent This work was financially supported by the Czech Science Foundation, Project No. 20-14547S.
\end{acknowledgments}


\begin{thebibliography}{99}

\bibitem{schick}
M. Schick, in {\it Liquids and Interfaces}, edited by J. Chorvolin, J. F. Joanny, and J. Zinn-Justin (Elsevier, New York, 1990).

\bibitem{dietrich}
S. Dietrich, in {\it Phase Transitions and Critical Phenomena}, edited by C. Domb and J. L. Lebowitz (Academic, New York, 1988), Vol. 12.

\bibitem{sullivan}
D. E. Sullivan and M. M. Telo da Gama, in {\it Fluid Interfacial Phenomena}, edited by C. A. Croxton (Wiley, New York, 1985).

\bibitem{evans90}
R. Evans, J. Phys.: Condens. Matter {\bf 2}, 8989 (1990).

\bibitem{hauge}
E. H. Hauge, Phys. Rev. A {\bf 46}, 4994 (1992).

\bibitem{rejmer}
K. Rejmer, S. Dietrich, and M. Napirk\'owski, Phys. Rev. E {\bf 60}, 4027 (1999).

\bibitem{wood99}
A. O. Parry, C. Rasc\'on, and A. J. Wood, Phys. Rev. Lett. {\bf 83}, 5535 (1999).

\bibitem{abraham02}
D. B. Abraham and Macio\l ek, Phys. Rev. Lett. {\bf 89}, 286101 (2002).

\bibitem{delfino}
G. Delfino and A. Squarcini, Phys. Rev. Lett. {\bf 113}, 066101 (2014).

\bibitem{binder03}
A. Milchev, M. M\"{u}ller, K. Binder, and D. P. Landau, Phys. Rev. Lett{ \bf 90}, 136101 (2003); Phys. Rev. E {\bf 68}, 031601 (2003).

\bibitem{bernardino}
N. R. Bernardino, A. O. Parry, and J. M. Romero-Enrique, J. Phys.: Condens. Matter {\bf 24}, 182202 (2011).

\bibitem{our_prl}
A. Malijevsk\'y and A. O. Parry, Phys. Rev. Lett. {\bf 110}, 166101 (2013).

\bibitem{our_wedge}
A. Malijevsk\'y and A. O. Parry, J. Phys.: Condens. Matter {\bf 25}, 305005 (2013).

\bibitem{darbellay}
G. A. Darbellay and J. M. Yeomans, J. Phys. A {\bf 25}, 4275 (1992).

\bibitem{evans_cc}
C. Rasc\'on, A. O. Parry, N. B. Wilding, and R. Evans, Phys. Rev. Lett. {\bf 98}, 226101 (2007).

\bibitem{tasin}
M. Tasinkevych and S. Dietrich, Eur. Phys. J. {\bf E23}, 117 (2007).

\bibitem{mistura}
L. Bruschi and G. Mistura, J. Low Temp. Phys. {\bf 157}, 206 (2009).

\bibitem{hofmann}
T. Hofmann, M. Tasinkevych, A. Checco, E. Dobisz, S. Dietrich, and B. M. Ocko, Phys. Rev, Lett. {\bf 104}, 106102 (2010).

\bibitem{schoen}
H. Boelen, A. O. Parry, E. Diaz-Herrera and M. Schoen, Eur. Phys. J. E {\bf 25}, 103 (2008).

\bibitem{mal_groove}
A. Malijevsk\'y, J. Chem. Phys. {\bf 137}, 214704 (2012).

\bibitem{parry_groove}
C. Rasc\'on, A. O. Parry, R. N\"{u}rnberg, A. Pozzato, M. Tormen, L Bruschi, and G. Mistura, J. Phys.: Condens. Matter {\bf 25}, 192101 (2013).

\bibitem{mistura13}
G. Mistura, A. Pozzato, G. Grenci, L. Bruschi, and M. Tormen, Nat. Commun. {\bf 4}, 2966 (2013).

\bibitem{our_groove}
A. Malijevsk\'y and A. O. Parry,  J. Phys: Condens. Matter {\bf 26},  355003  (2014).

\bibitem{monson}
D. Schneider, R. Valiullin, and P. A. Monsosn, Langmuir {\bf 30}, 1290 (2014).

\bibitem{fan}
C. Fan, D.D. Do, and D. Nicholson, Mol. Simul. {\bf 41}, 245 (2014).

\bibitem{het_groove}
 A. O. Parry, A. Malijevsk\'y and C.Rasc\'on, Phys. Rev. Lett. {\bf 113},  146101  (2014).

\bibitem{bruschi2}
L. Bruschi, G. Mistura, P.T.M. Nguyen, D.D. Do, D. Nicholson, S. J. Park, and W. Lee, Nanoscale {\bf 7}, 2587 (2015).

\bibitem{fin_groove_prl}
A. Malijevsk\'y and A. Parry, Phys. Rev. Lett. {\bf 120}, 135701 (2018).


\bibitem{nature}
C. Rasc\'on  and A. O. Parry, Nature {\bf 407}, 6807 (2000).

\bibitem{carlos}
C. Rasc\'on, Phys. Rev. Lett. {\bf 98}, 199801 (2007).

\bibitem{exp1}
L. Bruschi, A. Carlin, and G. Mistura, Phys. Rev. Lett. {\bf 89}, 166101 (2002).

\bibitem{exp2}
O. Gang, K. J. Alvine, M. Fukuto, P.S. Pershan and C. T. Black,  Phys. Rev. Lett. {\bf 95}, 217801 (2005).

\bibitem{osc}
M. Posp\'\i\v sil, A. O. Parry, and A. Malijevsk\'y, Phys. Rev. E {\bf 105}, 064801 (2022).


\bibitem{lipowsky84}
R. Lipowsky, Phys. Rev. Lett. {\bf 52}, 1429 (1984).

\bibitem{lipowsky85}
R. Lipowsky, Phys. Rev. B {\bf 32}, 1731 (1985).

\bibitem{fisher}
M. E. Fisher, J. Chem. Soc. Faraday Trans. 2  {\bf 82}, 1569 (1986).

\bibitem{lipowsky87}
R. Lipowsky and M. E. Fisher, Phys. Rev. B {\bf 36}, 2126 (1987).

\bibitem{abraham}
D. B. Abraham and D. A. Huse, Phys. Rev. B {\bf 38}, 7169 (1998).


\bibitem{row}
J. S. Rowlinson and B. Widom, {\it Molecular Theory of Capillarity} (Oxford: Clarendon, 1989).

\bibitem{hend}
J. R. Henderson in {\it Fundamentals of Inhomoheneous Fluids}, ed. by D. Henderson, Marcel Dekker, New York (1992).

\bibitem{evans}
R. Evans, in {\it Liquids and Interfaces}, edited by J. Chorvolin, J. F. Joanny, and J. Zinn-Justin (Elsevier, New York, 1990).

\bibitem{abraham86}
D. B. Abraham and E. R. Smith, J. Stat. Phys. {\bf 43}, 621 (1986).

\bibitem{md1}
A. Malijevsk\'y and A. O. Parry, Phys. Rev Lett. {\bf 127}, 115703 (2021).

\bibitem{md2}
A. Malijevsk\'y and A. O. Parry, Phys. Rev E {\bf 104}, 044801 (2021).

\bibitem{andelman}
D. Andelman, J.-F. Joanny, and M. O. Robbins, Europ. Phys. Lett. {\bf 7}, 731 (1988).

\bibitem{bieker}
M. P. Gelfand and R. Lipowsky, Phys. Rev. B {\bf 36}, 8725 (1987).

\bibitem{bieker}
T. Bieker and S. Dietrich, Physica A {\bf 252}, 85 (1998).

\bibitem{stewart}
M. C. Stewart and R. Evans, Phys. Rev. E {\bf 71}, 011602 (2005).

\bibitem{morgan}
A. O. Parry, C. Rasc\'on, and L. Morgan, J. Chem. Phys.  {\bf 7124}, 151101 (2006).

\bibitem{nold}
 A. Nold, A. Malijevsk\'y, and S. Kalliadasis, Phys. Rev. E {\bf 84}, 021603 (2011).

\bibitem{parry02}
A. O. Parry,  M. J. Greenall, and A. J. Wood, J. Phys. : Condens. Matter {\bf 14}, 1169 (2002).

\bibitem{evans79}
R. Evans, Adv. Phys. {\bf 28}, 143 (1979).

\bibitem{ros}
Y. Rosenfeld,  Phys. Rev. Lett. {\bf 63}, 980 (1989).

\bibitem{parry91}
A. O. Parry, J. Phys. A: Math. Gen. {\bf 24}, 1335 (1991).

\bibitem{tas}
M. Tasinkevych and S. Dietrich, Eur. Phys. J. E {\bf 23}, 117 (2007).






\end{thebibliography}
\end{document}